\documentclass[12pt]{article}
\usepackage{multicol}
\usepackage{amssymb}
\usepackage{amsfonts,amssymb,amsmath,amsthm}
\usepackage{epsfig}
\usepackage{slashed}
\usepackage{graphicx}
\usepackage{subfig}
\usepackage{mathrsfs}
\usepackage{bbm}
\usepackage{cite}
\usepackage{enumerate}
\usepackage{slashbox}
\usepackage{color}
\usepackage[colorlinks,linkcolor=red,anchorcolor=red,citecolor=green]{hyperref}
\makeatletter % `@' now normal "letter"
\@addtoreset{equation}{section}
\makeatother  % `@' is restored as "non-letter"

\makeatletter

\newcommand{\Rmnum}[1]{\expandafter\@slowromancap\romannumeral #1@}
\makeatother

\begin{document}
\renewcommand{\thefootnote}{\fnsymbol{footnote}}
%\begin{titlepage}

\vspace{10mm}
\begin{center}
{\Large\bf Hawking Radiation of Five-dimensional Charged Black Holes with Scalar Fields}
\vspace{16mm}

{{\large Yan-Gang Miao${}^{}$\footnote{\em E-mail: miaoyg@nankai.edu.cn}
and Zhen-Ming Xu}${}^{}$\footnote{\em E-mail: xuzhenm@mail.nankai.edu.cn}

\vspace{6mm}
${}^{}${\normalsize \em School of Physics, Nankai University, Tianjin 300071, China}

}

\end{center}

\vspace{10mm}
\centerline{{\bf{Abstract}}}
\vspace{6mm}
We investigate the Hawking radiation cascade from the five-dimensional charged black hole with a scalar field coupled to higher-order Euler densities in a conformally invariant manner. We give the semi-analytic calculation of greybody factors for the Hawking radiation. Our analysis shows that the Hawking radiation cascade from this five-dimensional black hole is extremely sparse. The charge enhances  the sparsity of the Hawking radiation, while the conformally coupled scalar field reduces this sparsity.
\vskip 20pt
\noindent
{\bf PACS Number(s)}: 04.70.-s, 04.70.Bw, 04.70.Dy


%\end{titlepage}

%\newpage
\renewcommand{\thefootnote}{\arabic{footnote}}
\setcounter{footnote}{0}
\setcounter{page}{2}
\pagenumbering{arabic}
%\tableofcontents
\vspace{1cm}

\section{Introduction}
Black holes, as a fascinating and elegant object, are increasingly popular in classical and quantum gravity theories. In particular, in virtue of semi-classical quantum field theory, Hawking~\cite{SH} revealed that black holes are actually characterized by thermally distributed emission spectra, not completely black. This intriguing phenomenon, known as the Hawking radiation, is now considered as certainly one of the theoretical topics of primary importance in fundamental physics. Meanwhile, a plenty of works~\cite{DP1,DP2,PW,ETM,HV,HK,CCG,SNS,HMR,PNW,TPP,PKP,SH2,FS1,FS2,KKZ,KZ,CAAB} devote to calculating the Hawking radiation in order to explore a convincing insight into a possible quantum gravity theory.

In this paper, we make some efforts to discuss the Hawking radiation for a charged black hole associated with a scalar field coupled to higher-order Euler densities in a conformally invariant manner in a five-dimensional spacetime~\cite{GGO,GGGO,HM,YZ}. In the situation of the scalar field conformally coupling to the Einstein gravity, the well-known no-hair theorem forbids~\cite{BT,CK} the existence of scalar hairy black holes, especially in $D>4$ dimensions. However, the recent research shows~\cite{OR,GLOR,MC} that the existence of scalar hairy black holes is becoming possible in arbitrary $D>4$ dimensions when a scalar field couples conformally to higher-order Euler densities. This gravity theory is the most general scalar field/gravity coupling theory whose field equations are of second order for both gravity and matter, and it can be regarded as a generalization of the Horndeski theory~\cite{GWH,YGMX1}. Hence, it is meaningful and necessary in this sense to study a five-dimensional black hole as a banner of high dimensional black holes with a scalar field conformally coupled to higher-order Euler densities. Here we mainly focus on the investigation of the sparsity of the Hawking radiation flow in order to capture the physical picture of the Hawking radiation out of the five-dimensional black hole with a scalar field. Taking a specific situation (see eq.~(\ref{action})) in ref.~\cite{GLOR} as our starting point, we calculate semi-analytically the Hawking radiation, and our result indicates that the Hawking radiation cascade from this five-dimensional charged black hole with a scalar field is extremely sparse. Specifically, the charge enhances the sparsity of the Hawking radiation, while the conformally coupled scalar field reduces this sparsity.

The paper is organized as follows. In section \ref{sec2}, we give a brief review of the analytic solution of the charged black hole with a scalar field in five dimensions. In section \ref{sec3}, we calculate the total Hawking radiation power and the greybody factor for this five-dimensional black hole. We then analyze the sparsity of the Hawking radiation cascade in section \ref{sec4}. Finally, we devote to a brief  summary in section \ref{sec5}. Note that the geometric units, $\hbar=c=k_{_B}=G=1$, are adopted throughout this paper.

\section{A Brief Review of the Charged Black Hole with a Scalar Field in Five Dimensions}\label{sec2}
We proceed to reviewing the five-dimensional charged black hole associated with a scalar field coupled to higher-order Euler densities in a conformally invariant manner. As a specific situation of the literature~\cite{GGGO,GLOR}, the model's action reads\footnote{In fact, we take $a_{k\neq 1}=0$ and $b_{k=0,1,2}\neq 0$ in ref.~\cite{GLOR}.}
\begin{equation}
\emph{I}=\frac{1}{\kappa}\int \text{d}^5 x\sqrt{-g}\left(R-\frac14 F^2+\kappa \mathscr{L}_{m}\right), \label{action}
\end{equation}
where $\kappa=16\pi$, $g=\text{det}(g_{\mu\nu})$, $R$ is the scalar curvature, $F$ the electromagnetic field strength, and $g_{\mu\nu}$ the metric with mostly plus signatures. Accordingly, the matter Lagrangian $\mathscr{L}_{m}$ takes the following form,
\begin{eqnarray}
\mathscr{L}_{m}= b_0 \phi^{15}+b_1 \phi^7 {S_{\mu\nu}}^{\mu\nu}+b_2 \phi^{-1}\left({S_{\mu\lambda}}^{\mu\lambda}{S_{\nu\delta}}^{\nu\delta}
-4{S_{\mu\lambda}}^{\nu\lambda}{S_{\nu\delta}}^{\mu\delta}+{S_{\mu\nu}}^{\lambda\delta}{S^{\nu\mu}}_{\lambda\delta}\right), \label{lag}
\end{eqnarray}
where $b_0$, $b_1$, and $b_2$ are coupling constants and ${S_{\mu\nu}}^{\lambda\delta}$ is a four-rank tensor.

The above model admits an exactly charged black hole solution in five dimensions~\cite{HM,GGGO},
\begin{equation}\label{metric}
\text{d}s^2=-f(r) \text{d}t^2+\frac{\text{d}r^2}{f(r)}+r^2 (\text{d}\theta_1^2+\sin^2 \theta_1\text{d}\theta_2^2+\sin^2 \theta_1\sin^2 \theta_2\text{d}\varphi^2),
\end{equation}
where $0\leq \theta_i<\pi$ $(i=1,2)$, $0\leq \varphi<2\pi$, and the function $f(r)$ takes the form,
\begin{equation}
f(r)=1-\frac{m}{r^2}-\frac{q}{r^3}+\frac{e^2}{r^4}, \label{lapfun}
\end{equation}
with integration constants $m$ and $e$ related to the mass and electric charge of this black hole, respectively. In addition, the constant $q$ of the scalar field $\phi$ conformally coupling to higher-order Euler densities has the form,
\begin{equation}
q=\frac{64\pi}{5}\varepsilon b_1\left(-\frac{18b_1}{5b_0}\right)^{3/2}, \label{defq}
\end{equation}
with $\varepsilon=-1,0,1$. Among the three coupling constants, $b_0$, $b_1$, and $b_2$, there exists an additional constraint: $10b_0 b_2=9b_1^2$, to ensure the formation of this black hole. The scalar field configuration and the Maxwell gauge potential take the forms,
\begin{eqnarray}
\phi(r)&=&\frac{\varepsilon}{r^{1/3}}\left(-\frac{18b_1}{5b_0}\right)^{1/6},\\
 A&=&\frac{\sqrt{3}e}{r^2}.
\end{eqnarray}

The black hole mass $M$ and the temperature $T$ are given by~\cite{GGGO,HM,YZ}
\begin{eqnarray}
M&=&\frac{3\pi}{8}m=\frac{3\pi}{8}\left(r_h^2-\frac{q}{r_h}+\frac{e^2}{r_h^2}\right), \label{enth}\\
T&=&\frac{1}{2\pi r_h}+\frac{q}{4\pi r_h^4}-\frac{e^2}{2\pi r_h^5}, \label{char}
\end{eqnarray}
where $r_h$ is the horizon radius of the black hole that is taken to be the largest real positive root of $f(r)=0$. In the following sections, we shall investigate the Hawking radiation cascade from this five-dimensional conformally coupled scalar field black hole.

\section{Hawking Radiation and Greybody Factor}\label{sec3}
We start with the emission of neutral scalar particles. Now we consider the power spectrum, i.e. the energy emitted per unit time for this five-dimensional charged black hole~\cite{FSAM},
\begin{equation}
\frac{\text{d}E(\omega)}{\text{d}t}=\sum_l \sigma_l(\omega)\frac{\omega}{e^{\omega/T}-1}\frac{\text{d}^4\textbf{k} \text{d}A}{(2\pi)^4},\label{dpower}
\end{equation}
where $T$ is temperature, $\text{d}A$ is infinitesimal surface area, and $\sigma_l(\omega)$ are dimensionless greybody factors which depend on the angular momentum quantum number $l$ and frequency $\omega$ of emitted scalar particles. For massless particles, $|\textbf{k}|=\omega$ and the phase space integration reduces to the one over emitted frequency $\omega$. Hence, the total Hawking radiation power in the finite surface area $A$ is
\begin{equation}
P=\frac{A}{8\pi^2}\sum_l \int_0^{\infty}\sigma_l(\omega)\frac{\omega^4 \text{d}\omega}{e^{\omega/T}-1}. \label{power}
\end{equation}
%For this finite surface area $A$, we make a note. 
The effective surface area of four-dimensional Schwarzschild black holes is $27/4$ times the area of the event horizon in order to smoothly match the low and high frequency results~\cite{DP1}. Here for the five-dimensional black hole, we can still approximatively regard its finite surface area $A$ as the horizon area in the present paper, i.e. $A=2\pi^2 r_h^3$, and such an estimation, in general, does not affect the following analysis about the Hawking radiation cascade qualitatively.

The most important content of the Hawking radiation, more precisely, the semi-classical Hawking radiation, is the calculation of greybody factors, which can modulate emission probabilities of Hawking gravitational quanta. The surrounding of the black hole, from the horizon radius to infinity, can be taken as an effective barrier. The greybody factor is the transmission probability of these gravitational quanta through this effective barrier. The relevant behaviors can be governed by the Klein-Gordon equation. For simplicity, we now discuss the excitations for a massless scalar field $\Phi$,\footnote{We emphasize that this emitted massless scalar field is different from the conformally coupled scalar field.}
\begin{equation}
\frac{1}{\sqrt{-g}}\partial_{\mu}(\sqrt{-g}g^{\mu\nu}\partial_{\nu}\Phi)=0. \label{kgeq}
\end{equation}
This field can be written~\cite{HK,CCG,SNS} by virtue of the separation of variables as follows,
\begin{equation}
\Phi(t,r,\theta_1,\theta_2,\varphi)=e^{-i\omega t}U(r)Y(\theta_1,\theta_2,\varphi). \label{sf}
\end{equation}
Inserting eqs.~(\ref{metric}) and~(\ref{sf}) into eq.~(\ref{kgeq}), and completing the separation of variables, we can obtain the radial equation,
\begin{equation}\label{radeq}
\frac{f}{r^3}\frac{\text{d}}{\text{d}r}\left(r^3 f\frac{\text{d}U(r)}{\text{d}r}\right)+\left(\omega^2-f \cdot \frac{l(l+2)}{r^2}\right)U(r)=0,
\end{equation}
where $f$ is described by eq.~(\ref{lapfun}) and $l(l+2)$ comes from the contribution of the angular equation $Y(\theta_1,\theta_2,\varphi)$. Then, introducing the so-called `tortoise' radial coordinate, defined by $\text{d}r_{\star} \equiv \text{d}r/f$,
and the variable transformation,
\begin{equation}
U(r)=\frac{R(r)}{r^{3/2}},
\end{equation}
we reduce the radial equation~(\ref{radeq}) to a Schr\"{o}dinger-like one,
\begin{equation}\label{redradeq}
\left(\frac{\text{d}^2}{\text{d}r_{\star}^2}+\omega^2-V_{\text{eff}}\right)R(r)=0,
\end{equation}
where the effective curvature potential takes the form,
\begin{equation}\label{veff}
V_{\text{eff}}=f\left\{\frac{l(l+2)}{r^2}+\frac{3f}{4r^2}+\frac{3f^{\prime}}{2r}\right\}.
\end{equation}
For the near horizon region $r\rightarrow r_h$, we can see $r_{\star}\rightarrow -\infty$ and $V_{\text{eff}}\rightarrow 0$. For the far-field region $r\rightarrow\infty$, we have $r_{\star}\rightarrow +\infty$ and $V_{\text{eff}}\rightarrow 0$. Hence, the radial solution only contains the ingoing mode $e^{-i\omega r_{\star}}$ near the horizon region, while in the far-field region, it comprises both the ingoing mode $e^{-i\omega r_{\star}}$ and outgoing mode  $e^{i\omega r_{\star}}$. As a result, the greybody factor can be obtained by calculating the absorption coefficient of such a scattering problem, but unfortunately as far as we know, there are few exactly analytical results for the greybody factor. Therefore, a wide variety of techniques have been used to deal with the estimate of black hole greybody factors, like as the extremal limit and asymptotically high or low frequencies limit~\cite{HK,CCG,SNS,HMR,PNW,TPP,PKP,MCFL,MCFL2}. Interestingly, the literatures~\cite{BMV,BNMV,BNMV2,BCNMV} give a hopefully simple analytic bound on the greybody factor. Here we adopt this approach to tackle our problem. The general semi-analytic bounds for greybody factors read as
\begin{equation}\label{boundgf}
\sigma_l(\omega) \geq \text{sech}^2 \left(\int_{-\infty}^{+\infty}\vartheta \text{d}r_{\star}\right),
\end{equation}
where
\begin{equation}
\vartheta=\frac{\sqrt{(h^{\prime})^2+(\omega^2-V_{\text{eff}}-h^2)^2}}{2h},
\end{equation}
and $h$ is a certain positive function which only satisfies the conditions $h(r_{\star})>0$ and $h(-\infty)=h(+\infty)=\omega$. Here we just take a simple form $h=\omega$, resulting in that the integration of eq.~(\ref{boundgf}) reduces to be
\begin{equation}\label{reducedint}
\int_{-\infty}^{+\infty}\vartheta \text{d}r_{\star}=\int_{r_h}^{+\infty}\frac{V_{\text{eff}}}{2\omega f}\text{d}r.
\end{equation}
With the aid of eqs.~(\ref{lapfun}),~(\ref{enth}),~(\ref{veff}),~(\ref{boundgf}), and~(\ref{reducedint}), we finally obtain the analytic bounds to the greybody factor,
\begin{eqnarray}\label{gbf}
\sigma_l(\omega)\geq \text{sech}^2 \left\{\frac{l(l+2)}{2\omega}\frac{1}{r_h}
+\frac{1}{2\omega}\left(\frac{3}{2r_h}+\frac{3q}{16r_h^4}-\frac{3e^2}{10r_h^5}\right)\right\}.
\end{eqnarray}

\begin{figure}
\begin{center}
  \begin{tabular}{cc}
    \includegraphics[width=75mm]{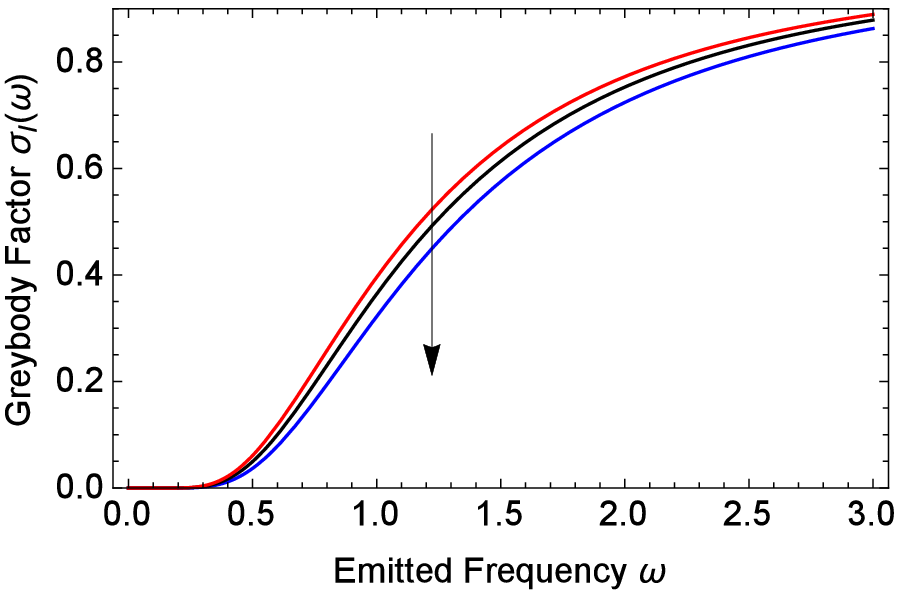} &
    \includegraphics[width=75mm]{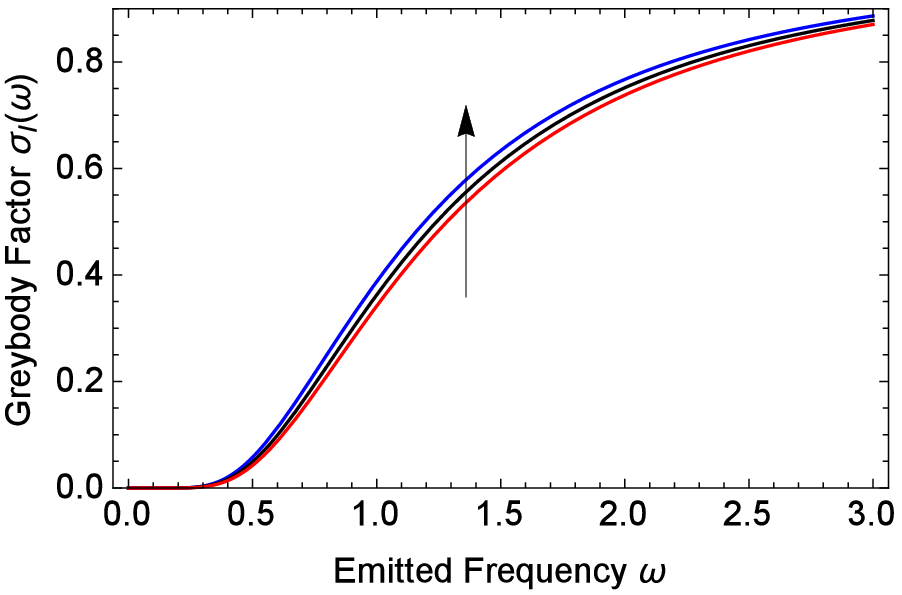} \\
  \end{tabular}
\end{center}
\caption{(\textbf{Left}) Plot of the greybody factor $\sigma_l(\omega)$ with respect to frequency $\omega$ at $l=1$, $e=1$, and $r_h=2$ for $q=-14, -5$, and $8$, respectively. The arrow represents the direction of the increasing parameter $q$. (\textbf{Right}) Plot of the greybody factor $\sigma_l(\omega)$ with respect to frequency $\omega$ at $l=1$, $q=1$, and $r_h=2$ for $e=0, 2.8$, and $4.1$, respectively. The arrow represents the direction of the increasing parameter $e$.}
\label{geryf}
\end{figure}

\begin{figure}
\begin{center}
  \begin{tabular}{ccc}
    \includegraphics[width=51mm]{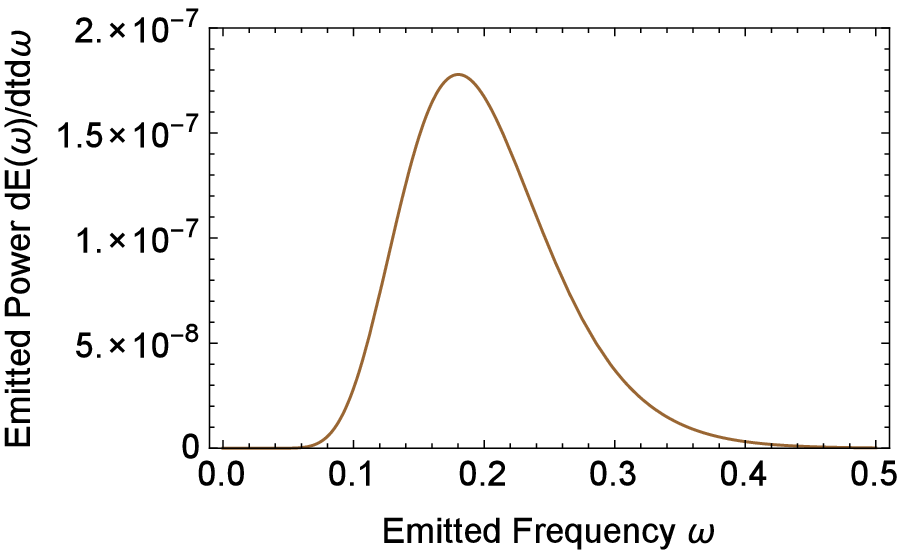} &
    \includegraphics[width=51mm]{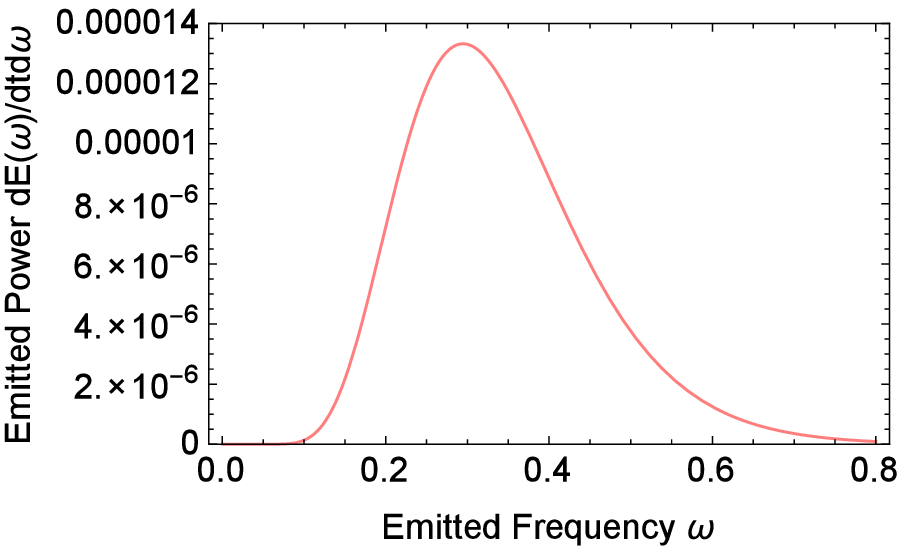} &
    \includegraphics[width=51mm]{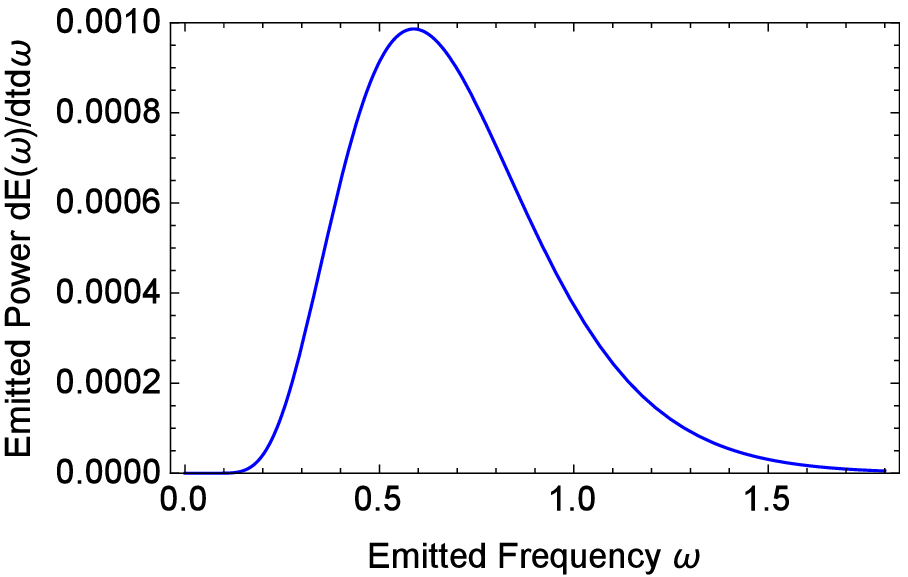} \\
    \includegraphics[width=51mm]{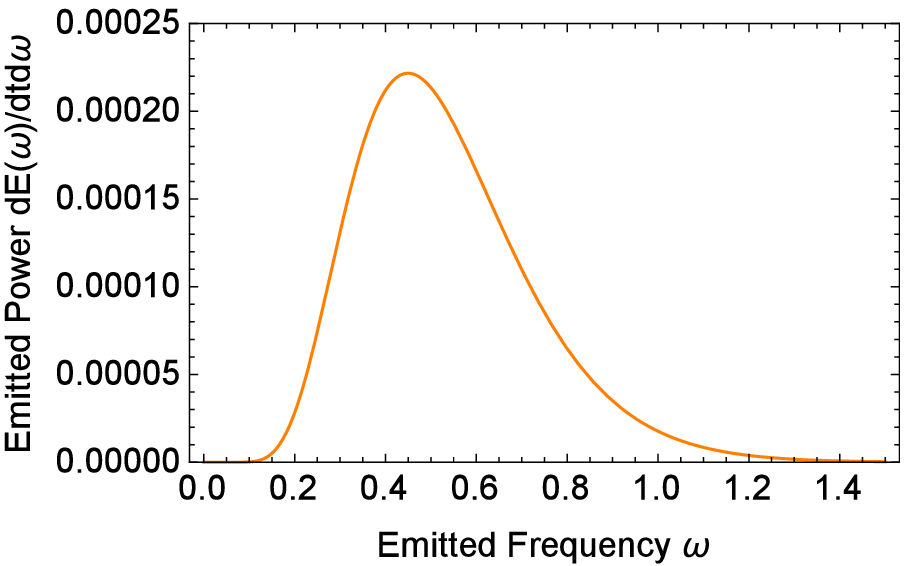} &
    \includegraphics[width=51mm]{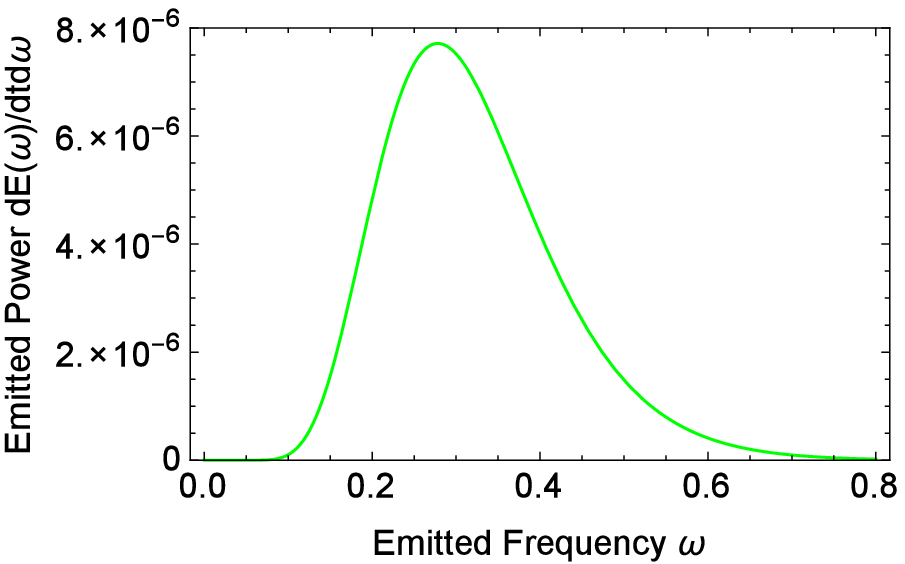} &
    \includegraphics[width=51mm]{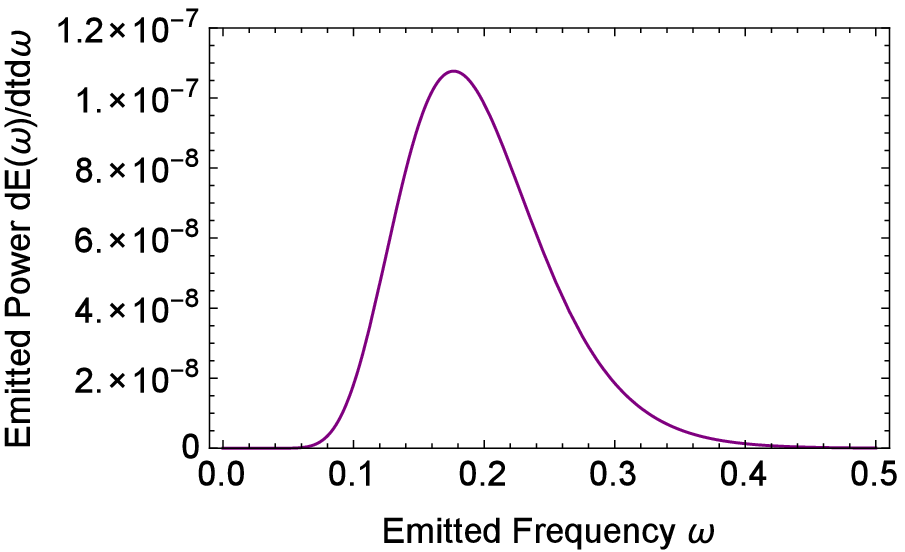} \\
  \end{tabular}
\end{center}
\caption{(\textbf{Top}) Plots of the power spectrum $\text{d}E(\omega)/(\text{d}t \text{d}\omega)$ with respect to frequency $\omega$ at $l=1$, $e=1$, and $r_h=2$ for $q=-10, -5$, and $8$, respectively,  from left to right. (\textbf{Down}) Plots of the power spectrum $\text{d}E(\omega)/(\text{d}t \text{d}\omega)$ with respect to frequency $\omega$ at $l=1$, $q=1$, and $r_h=2$ for $e=0, 2.8$, and $3.5$, respectively,  from left to right.}
\label{powerf}
\end{figure}

In accordance with the above result, we can predict that the parameters $q$ and $e$ will play an opposite role in the effects to the greybody factor due to the minus sign in front of $e^2$. In order to explain this phenomenon more intuitively, we plot the greybody factor (eq.~(\ref{gbf})) and the power spectrum (eq.~(\ref{dpower})) in Figures \ref{geryf} and \ref{powerf}, respectively, for this five-dimensional black hole. We know that the greybody factor equals one for the perfect black body radiation, but it is less than one for the Hawking radiation of black holes. From Figure \ref{geryf}, we can see that for given $l$, $e$, and $r_h$, the value of $q$ depresses the greybody factor when it becomes large, while the value of $e$ enhances the greybody factor when it is becomes large.\footnote{Note that the range of $q$ is from the negative infinity to the positive infinity, while that of the parameter $e^2$ is just from zero to the positive infinity.} From Figure \ref{powerf}, we can clearly observe that the peak power spectrum is gradually increasing with the increasing of $q$. On the contrary, with the increasing of $e$, the peak power spectrum is gradually decreasing. In a word, the parameters $q$ and $e$ play an opposite role in the effects to the greybody factor and the power spectrum.

\section{Sparsity of Hawking Radiation}\label{sec4}
In order to describe the semi-classical Hawking radiation flux out of the black hole, the literatures~\cite{FSAM,Hod,Hod2} have introduced a dimensionless ratio to characterize the feature of the Hawking cascade. Now we briefly review this strategy. The dimensionless ratio is defined as
\begin{equation}\label{sphw}
\eta=\frac{\tau_{\text{gap}}}{\tau_{\text{emission}}},
\end{equation}
where $\tau_{\text{gap}}$ is the characteristic time interval between the emissions of successive Hawking gravitational quanta out of the black hole and $\tau_{\text{emission}}$ is the characteristic timescale required for an individual Hawking gravitational quantum to be emitted from the black hole. They are depicted in detail by the following characteristics.
\begin{itemize}
  \item $\tau_{\text{emission}}$ can be bounded by the time-period, which can be considered as the time the corresponding emitted wave field needs to  complete a full oscillation cycle. At the peak frequency $\omega_{\text{peak}}$ which occurs when $\omega^4 /(e^{\omega/T}-1)$ is maximized, the quanta emitted can only be temporally localized to within a few oscillation periods. Hence,
      \begin{equation}
      \tau_{\text{emission}}\geq \frac{2\pi}{\omega_{\text{peak}}}=\frac{2\pi}{T[4+W(-4e^{-4})]},\label{emission}
      \end{equation}
      where $W(x)$ is the Lambert-W function.
  \item $\tau_{\text{gap}}$ can be defined by using the reciprocal of the total Hawking radiation power $P$ (eq.~(\ref{power})) and the characteristic peak frequency $\omega_{\text{peak}}$. That is,
      \begin{equation}
      \tau_{\text{gap}}=\frac{\omega_{\text{peak}}}{P}=\frac{T[4+W(-4e^{-4})]}{P}.\label{gap}
      \end{equation}
  \item $\eta\gg 1$ implies that the Hawking cascade is extremely sparse. The larger the ratio $\eta$ is, the sparser the Hawking radiation flow is.
  \item $\eta\ll 1$ implies that the Hawking cascade is almost continuous. The smaller the ratio $\eta$ is, the more continuous the Hawking radiation flow is.
\end{itemize}

Next we use this strategy to analyze the five-dimensional black hole. Numerical results of the dimensionless ratio $\eta$ can be seen in Table \ref{biao1} and Table \ref{biao2}. We can observe that the dimensionless ratio $\eta \gg 1$ for the five-dimensional black hole (regardless of the cases of $q=10$, $20$, and $40$ in Table \ref{biao1}, in which the mass of the black hole is negative and the related analyses will be given in the next section), which implies that the Hawking cascade is extremely sparse, i.e. the radiation of the five-dimensional black hole typically emits one massless quantum at a time. More importantly, when the coupled parameter $q$ becomes small, the dimensionless ratio $\eta$ will become large enough, causing that the Hawking radiation flow is extremely sparse as shown in Table \ref{biao1}. While from Table \ref{biao2}, we can obtain that the larger the charge parameter $e$ is, the sparser the Hawking radiation flow is. Hence, the parameters $q$ and $e$ play an opposite role in the effects to the Hawking radiation flow. The charge enhances the sparsity of the Hawking radiation, while the conformally coupled scalar field reduces this sparsity under the condition that the parameters $q$ and $e$ are getting large.

\begin{table}[!htp]
\begin{center}
\begin{tabular}{c|*{4}{c}|ccc}
\hline
\multicolumn{7}{c} {The dimensionless ratio $\eta$ at $l=1$, $e=1$ and $r_h=2$} \\ \hline \hline
$q$       & -10     & -5       & 2        & 6       & 10       & 20      & 40 \\ \hline
$\eta$    & 6149.5  & 1730.2   & 176.72   & 75.958  & 38.596   & 10.908  & 2.1639\\ \hline
\end{tabular}
\end{center}
\caption{The numerical results of the dimensionless ratio $\eta$ for different $q$ values.}
\label{biao1}
\end{table}

\begin{table}[!htp]
\begin{center}
\begin{tabular}{c|*{7}{c}}
\hline
\multicolumn{7}{c} {The dimensionless ratio $\eta$ at $l=1$, $q=1$ and $r_h=2$} \\ \hline \hline
$e$      & 0       & 0.9     & 1.5    & 2.0    & 2.4    & 2.8     & 3.4      \\ \hline
$\eta$   & 176.08  & 215.55  & 318.79 & 548.91 & 1048.2 & 2724.6  & 42519  \\ \hline
\end{tabular}
\end{center}
\caption{The numerical results of the dimensionless ratio $\eta$ for different $e$ values.}
\label{biao2}
\end{table}

\section{Summary}\label{sec5}
We give the semi-analytic calculation of the Hawking radiation for the five-dimensional black hole conformally coupled a scalar field. Our observation states clearly that the greybody factor always has the effect of `\textit{increasing}' the dimensionless ratio $\eta$ because it can `\textit{suppress}' the total Hawking radiation power $P$, and that the parameters $q$ and $e$ always play an opposite role in the effects to the Hawking radiation. This conclusion is mainly manifested in the following two aspects.
\begin{itemize}
  \item With the increasing of conformally coupled parameter $q$, the greybody factor is always decreasing, resulting in that the total Hawking radiation power $P$ increases and the dimensionless ratio $\eta$ decreases.
  \item With the increasing of charge parameter $e$, the greybody factor is always increasing, resulting in that the total Hawking radiation power $P$ decreases and the dimensionless ratio $\eta$ increases.
\end{itemize}

We observe that the total power spectrum (eq.~(\ref{power})), the greybody factor (eq.~(\ref{gbf})), and the sparsity of Hawking radiation (eq.~(\ref{sphw}))
depend  obviously on the black hole mass and the temperature. This characteristic is based on the two facts: i) the expression of black hole mass eq.~(\ref{enth}) is utilized in the calculation of the greybody factor, and ii) the power spectrum of Hawking radiation contains the temperature-dependent factor $1/(e^{\omega/T}-1)$, see eq.~(\ref{dpower}), and the sparsity depends on the temperature and the Hawking radiation power, see eqs.~(\ref{emission}) and~(\ref{gap}).
%the influence of temperature on the power spectrum of Hawking radiation is mainly reflected in the factor $1/(e^{\omega/T}-1)$, see . 
According to eqs.~(\ref{enth}) and~(\ref{char}), we see $M=M(r_h, q, e)$ and $T=T(r_h, q, e)$. For fixed $r_h$ and $q$, the black hole mass $M$ increases and the temperature $T$ decreases with the increasing of charge $e$. However, the black hole mass $M$ decreases and the temperature $T$ increases with the increasing of conformally coupled parameter $q$ for fixed $r_h$ and $e$. Furthermore, our analysis shows that the Hawking cascade from the five-dimensional black hole conformally coupled a scalar field  is extremely sparse. During the Hawking radiation, the black hole mass decreases rapidly and the temperature increases rapidly with the increasing of conformally coupled parameter $q$. In this situation, the radiation process is  intensified, i.e., the greybody factor decreases and the total Hawking radiation power increases, and thus the sparsity of the Hawking radiation reduces. On the contrary, the situation of the increasing of charge $e$ impedes the radiation process and thus enhances this sparsity.
Incidentally, there is a subtlety that the dimensionless ratio is approximately in the order of unit, $\eta\sim \text{O}(1)$, in the process of increasing the value of conformally coupled parameter $q$, which implies that the Hawking cascade from the five-dimensional conformally coupled scalar field black hole is becoming continuous as shown in Table \ref{biao1} for the cases of $q=10$, $20$, and $40$. Unfortunately, the mass of the black hole (see eq.~(\ref{enth})) in these cases is negative~\cite{HM}. With the increasing of $q$, the mass becomes negative and the dimensionless ratio takes $\eta \ll 1 $. However, the negative mass situation has also appeared in some Einstein-AdS gravity in the hyperbolic case, which can be related to the negative energy density matter~\cite{RBM,SRBM}. As a result, the Hawking cascade shows  a continuous characteristic if one accepts the negative mass black hole. This unwonted situation will be discussed in our future work.

\section*{Acknowledgments}
This work was supported in part by the National Natural Science Foundation of China under grant No.11675081. The authors would like to thank the anonymous referee for the helpful comments that indeed greatly improve this work.

\end{document}